\documentclass[prb,twocolumn]{revtex4}%

\usepackage{graphicx}
\usepackage{amsmath}
\usepackage{ulem}
\usepackage{color}

\newcommand{\be}{\begin{equation}}
\newcommand{\ee}{\end{equation}}
\newcommand{\bea}{\begin{eqnarray*}}
\newcommand{\eea}{\end{eqnarray*}}
\newcommand{\bean}{\begin{eqnarray}}
\newcommand{\eean}{\end{eqnarray}}

\begin{document}

\draft
\title
{Quantum interference and structure-dependent orbital-filling
effects on the thermoelectric properties of quantum dot molecules}
\author{Chih-Chieh Chen,$^{1,3}$ David M T Kuo$^{2}$ and Yia Chung Chang$^{3,4}$ }
\affiliation{$^{1}$Department of Physics, University of Illinois at
Urbana-Champaign, Urbana, Illinois 61801, USA}

\affiliation{$^{2}$Department of Electrical Engineering and
Department of Physics, National Central University, Chungli, 320
Taiwan}

\affiliation{$^{3}$Research Center for Applied Sciences, Academic
Sinica, Taipei, 11529 Taiwan} \affiliation{$^{4}$Department of
Physics, National Cheng-Kung University, Tainan, 70101 Taiwan}

\date{\today}

\begin{abstract}
{The quantum interference and orbital filling effects on the
thermoelectric (TE) properties of quantum dot molecules with high
figure of merit are illustrated via the full solution to the
Hubbard-Anderson model in the Coulomb blockade regime. It is found
that under certain condition in the triangular QD molecule (TQDM),
destructive quantum interference (QI) can occur, which leads to
vanishing small electrical conductance, while the Seebeck
coefficient is modified dramatically. When TQDM is in the charge
localization state due to QI, the Seebeck coefficient is seriously
suppressed at low temperature, but highly enhanced at high
temperature. Meanwhile, the behavior of Lorenz number reveals that
it is easier to block charge transport via destructive QI than the
electron heat transport at high temperatures. The maximum power
factor (PF) in TQDM occurs at full-filling condition. Nevertheless,
low-filling condition is preferred for getting maximum PF in
serially coupled triple QDs in general. In double QDs, the maximum
PF  can be achieved either with orbital-depletion or orbital-filling
as a result of electron-hole symmetry. Our theoretical work provides
a useful guideline for advancing the nanoscale TE technology.}
\end{abstract}

\maketitle
\section{Introduction}
To design solid-state coolers and power generators,[1-7] many
efforts seek efficient thermoelectric (TE) materials with the figure
of merit ($ZT$) larger than 3, which will lead to efficient TE
conversion, making the TE device competitive with conventional air
conditioners and power generators.[1,2] Furthermore, the energy
harvesting by using TE materials is one of several advanced
techniques considered for clean energies. The optimization of
$ZT=S^2G_eT/\kappa$ depends on the electrical conductance ($G_e$),
Seebeck coefficient (S), and thermal conductance ($\kappa$). $T$ is
the equilibrium temperature. These physical quantities are usually
related to one another. Mechanisms leading to the enhancement of
power factor ($PF=S^2G_e$) would also enhance the thermal
conductance. Consequently, it is difficult to obtain $ZT$ above one
in conventional bulk materials.[1]

Recently, quantum dot superlattice (QDSL) nanowires with impressive
$ZT$ values (larger than one) have been demonstrated
experimentally.[8] The power factor and thermal conductance become
independent thermoelectric variables under the condition
$\kappa_e/\kappa_{ph} \ll 1$, where $\kappa_e$ and $\kappa_{ph}$
denote, respectively, the electron thermal conductance and phonon
thermal conductance.[8] In the Coulomb blockade regime, electron
transport process is seriously suppressed by the electron Coulomb
interactions, causing $\kappa_e$ and $G_e$ to be reduced
simultaneously.[9] On the other hand, under the condition
$\kappa_e/\kappa_{ph} \ll 1$, one can increase the power factor (PF)
[10-15] and decrease the phonon thermal conductance [16-21]
simultaneously to optimize $ZT$. The ZT enhancement of QDSL mostly
arises from changes induced by dimensional confinement in the
electronic band structures as well as enhanced phonon scattering
resulting from the scattering at nanowire surface and interfaces
surrounding QDs.[1,2]

Although many theoretical efforts have investigated the PF
enhancement of QDs[10-15] and molecule junctions[22-24], quantum
interference (QI) and orbital filling effects on the PF optimization
are still puzzling due to the difficulty to treat many body effect
reliably in either QDSLs or molecules.[25] For example, in the
triangular QD molecule (TQDM), there are 923 electron correlation
functions appeared in the 4752 Green's functions need to be solved,
when electron Coulomb interactions are turned on, even though one
just considers one energy level for each QD. As a consequence, a
theoretical framework to treat adequately the many-body problem of a
molecular junction remains elusive. QI is a remarkable effect, which
influences the charge transport of QD array [6-8] and molecules
[22-24] with multiple quantum paths. When the energy levels of QDSLs
{or molecules are below the Fermi energy of electrodes, the orbital
filling effects can not be avoided for charge/heat transport. To
reveal the many-body effects on the TE properties of QD junction
system, we theoretically investigate  QD molecules (QDMs), including
double QD molecule (DQD), serially coupled triple QDs (SCTQD), and
TQDM with a full many-body solution which takes into account all
correlation functions resulting from electron Coulomb interactions
in QDMs. Our results indicate that the effects of QI and
structure-dependent orbital filling are quite significant in
determining the behaviors of TE coefficients for junctions of QDSL
or molecules ( such as benzene).

\section{Formalism}
Here we consider nanoscale semiconductor QDs, in which the energy
level separations are much larger than their on-site Coulomb
interactions and thermal energies. Thus, only one energy level for
each quantum dot needs to be considered. An extended Hubbard-Anderson model is employed to
simulate a QDMs connected to electrodes (see the inset of Fig. 1(b)).  The Hamiltonian of the
QDM junction is given by $H=H_0+H_{QD}$:

\begin{eqnarray}
H_0& = &\sum_{k,\sigma} \epsilon_k
a^{\dagger}_{k,\sigma}a_{k,\sigma}+ \sum_{k,\sigma} \epsilon_k
b^{\dagger}_{k,\sigma}b_{k,\sigma}\\ \nonumber &+&\sum_{k,\sigma}
V_{k,L}d^{\dagger}_{L,\sigma}a_{k,\sigma}
+\sum_{k,\sigma}V_{k,R}d^{\dagger}_{R,\sigma}b_{k,\sigma}+c.c
\end{eqnarray}
where the first two terms describe the free electron gas of left and
right electrodes. $a^{\dagger}_{k,\sigma}$
($b^{\dagger}_{k,\sigma}$) creates  an electron of momentum $k$ and
spin $\sigma$ with energy $\epsilon_k$ in the left (right)
electrode. $V_{k,\ell}$ ($\ell=L,R$) describes the coupling between
the electrodes and the left (right) QD. $d^{\dagger}_{\ell,\sigma}$
($d_{\ell,\sigma}$) creates (destroys) an electron in the $\ell$-th
dot.

\begin{small}
\begin{eqnarray}
H_{QD}&=& \sum_{\ell,\sigma} E_{\ell} n_{\ell,\sigma}+
\sum_{\ell} U_{\ell} n_{\ell,\sigma} n_{\ell,\bar\sigma}\\
\nonumber &+&\frac{1}{2}\sum_{\ell,j,\sigma,\sigma'}
U_{\ell,j}n_{\ell,\sigma}n_{j,\sigma'}
+\sum_{\ell,j,\sigma}t_{\ell,j} d^{\dagger}_{\ell,\sigma}
d_{j,\sigma},
\end{eqnarray}
\end{small}
where { $E_{\ell}$} is the spin-independent QD energy level, and
$n_{\ell,\sigma}=d^{\dagger}_{\ell,\sigma}d_{\ell,\sigma}$.
Notations $U_{\ell}$ and $U_{\ell,j}$ describe the intradot and
interdot Coulomb interactions, respectively. $t_{\ell,j}$ describes
the electron interdot hopping. Noting that the interdot Coulomb
interactions as well as intradot Coulomb interactions play a
significant role on the charge transport for semiconductor QDs [9]
and molecular structures.[22-24]

Using the Keldysh-Green's function technique,[26,27] the charge and
heat currents from reservoir $\alpha$ to the QDM junction are
calculated according to the Meir-Wingreen formula
\begin{eqnarray}
J_{\alpha}&=&\frac{ie}{h}\sum_{j\sigma}\int {d\epsilon}
\Gamma^\alpha_{j}(\epsilon) [ G^{<}_{j\sigma} (\epsilon)+ f_\alpha
(\epsilon)( G^{r}_{j\sigma}(\epsilon) \nonumber \\ &-&
G^{a}_{j\sigma}(\epsilon) ) ]\\ Q_{\alpha}
&=&\frac{i}{h}\sum_{j\sigma}\int {d\epsilon} (\epsilon-\mu_{\alpha})
\Gamma^\alpha_{j}(\epsilon) [ G^{<}_{j\sigma} (\epsilon) f_\alpha
(\epsilon) \nonumber \\ & &( G^{r}_{j\sigma}(\epsilon) -
G^{a}_{j\sigma}(\epsilon) ) ],
\end{eqnarray}
Notation $\Gamma^\alpha_{\ell}=\sum_k
|V_{k,\alpha,\ell}|^2\delta(\epsilon-\epsilon_k)$ is the tunneling
rate between the $\alpha$-th reservoir and the $\ell$-th QD.
$f_{\alpha}(\epsilon)=1/\{\exp[(\epsilon-\mu_{\alpha})/k_BT_{\alpha}]+1\}$
denotes the Fermi distribution function for the $\alpha$-th
electrode, where $\mu_\alpha$  and $T_{\alpha}$ are the chemical
potential and the temperature of the $\alpha$ electrode.
$\mu_L-\mu_R=\Delta V$ and $T_L-T_R=\Delta T$. $e$, $h$, and $k_B$
denote the electron charge, the Planck's constant, and the Boltzmann
constant, respectively. $G^{<}_{j\sigma} (\epsilon)$,
$G^{r}_{j\sigma}(\epsilon)$, and $G^{a}_{j\sigma}(\epsilon)$ are the
frequency domain representations of the one-particle lessor,
retarded, and advanced Green's functions
$G^{<}_{j\sigma}(t,t')=i\langle d_{j,\sigma}^\dagger (t')
d_{j,\sigma}(t) \rangle $, $G^{r}_{j\sigma}(t,t')=-i\theta
(t-t')\langle \{ d_{j,\sigma}(t),d_{j,\sigma}^\dagger (t') \}
\rangle $, and $G^{a}_{j\sigma}(t,t')=i\theta (t'-t)\langle \{
d_{j,\sigma}(t),d_{j,\sigma}^\dagger (t') \}  \rangle $,
respectively. {These one-particle Green's functions are related
recursively to other Green's functions and density-density
correlators via the few-body equation of motion,[25] which we solve
via an iterative numerical procedure to obtain all $n$-particle
Green's functions and correlators for the QDM.  The coupling between
electrons in the leads and electrons in QDM is included in the
self-energy term, $\Gamma^\alpha_{\ell}$. Because of this
approximation, our procedure is valid only in the Coulomb blockade
regime, but not the Kondo regime.[28]

Thermoelectric coefficients in the linear response regime are
\begin{eqnarray}
G_e&=& (\frac{\delta J_{\alpha}}{\delta \Delta V})_{\Delta T=0} \\
S&=& - (\frac{\delta J_{\alpha}}{\delta \Delta T})_{\Delta V=0}/ (\frac{\delta J_{\alpha}}{\delta \Delta V})_{\Delta T=0} \\
\kappa_e &=& (\frac{\delta Q_{\alpha}}{\delta \Delta T})_{\Delta
V=0}+(\frac{\delta Q_{\alpha}}{\delta \Delta V })_{\Delta T=0}S\\
&=&(\frac{\delta Q_{\alpha}}{\delta \Delta T})_{\Delta V=0}-S^2G_eT
\nonumber
\end{eqnarray}
where
\begin{eqnarray}
(\frac{\delta J_{\alpha}}{\delta \Delta V})_{\Delta T=0}&=&\frac{ie}{h}\sum_{j\sigma}\int {d\epsilon}\Gamma^\alpha_{j}(\epsilon)\times \\
&& [ \frac{\delta G^{<}_{j\sigma}(\epsilon)}{\delta
f_{\alpha}(\epsilon)} + ( G^{r}_{j\sigma}(\epsilon) \nonumber -
G^{a}_{j\sigma}(\epsilon) ) ]\frac{\delta f_{\alpha}(\epsilon)}{\delta \Delta V}  \\
(\frac{\delta J_{\alpha}}{\delta \Delta T})_{\Delta V=0}&=&\frac{ie}{h}\sum_{j\sigma}\int {d\epsilon}\Gamma^\alpha_{j}(\epsilon)\times \\
&& [ \frac{\delta G^{<}_{j\sigma}(\epsilon)}{\delta
f_{\alpha}(\epsilon)} + ( G^{r}_{j\sigma}(\epsilon) \nonumber -
G^{a}_{j\sigma}(\epsilon) ) ]\frac{\delta f_{\alpha}(\epsilon)}{\delta \Delta T}  \\
(\frac{\delta Q_{\alpha}}{\delta \Delta T})_{\Delta V=0}&=&\frac{i}{h}\sum_{j\sigma}\int {d\epsilon}\Gamma^\alpha_{j}(\epsilon)(\epsilon-E_F) \times \\
&& [ \frac{\delta G^{<}_{j\sigma} (\epsilon)}{\delta
f_{\alpha}(\epsilon)} + ( G^{r}_{j\sigma}(\epsilon) \nonumber -
G^{a}_{j\sigma}(\epsilon) ) ]\frac{\delta f_{\alpha}(\epsilon)
}{\delta
\Delta T} ,\\
(\frac{\delta Q_{\alpha}}{\delta \Delta V})_{\Delta T=0}&=&\frac{i}{h}\sum_{j\sigma}\int {d\epsilon}\Gamma^\alpha_{j}(\epsilon)(\epsilon-E_F) \times \\
&& [ \frac{\delta G^{<}_{j\sigma}(\epsilon)}{\delta
f_{\alpha}(\epsilon)} + ( G^{r}_{j\sigma}(\epsilon) \nonumber -
G^{a}_{j\sigma}(\epsilon) ) ] \frac{\delta
f_{\alpha}(\epsilon)}{\delta \Delta V}.
\end{eqnarray}
The quantity $\frac{\delta G^{(1)<}_{j\sigma}(\epsilon)}{\delta
f_{\alpha}(\epsilon)}$ is obtained by solving the variation of the
equation of motion with respect to the change in Fermi-Dirac
distribution, $f_{\alpha}(\epsilon)$. Here we have assumed the
variation of the correlation functions with respect to $\delta
f_{\alpha}(\epsilon)$ is of the second order. Note that we have to
take $\Delta V \rightarrow 0$ for the calculations of $(\frac{\delta
J_{\alpha}}{\delta \Delta V})_{\Delta T=0}$ and $(\frac{\delta
Q_{\alpha}}{\delta \Delta V})_{\Delta T=0}$. Meanwhile we taken
$\Delta T \rightarrow 0$ for Eqs.~(9) and (10). {$E_F$
is the Fermi energy of electrodes.} There is a Joule heating term
$\Delta V \times J_{\alpha}$ arising from Eq.~(4) which can be
ignored in the linear response regime.

It is a challenge to prove that whether the Eqs.~(3) and (4) can be
expressed in terms of transmission coefficient (or Landauer's
expression) when one takes into account all Green's functions and
correlation functions fully. To gain deeper insight into the
electron correlation effect, we adopt the approximation procedure of
Ref.~[9] and obtain  $G_e=e^2{\cal L}_{0}$, $S=-{\cal
L}_{1}/(eT{\cal L}_{0})$ and $\kappa_e=\frac{1}{T}({\cal
L}_{2}-{\cal L}^2_{1}/{\cal L}_{0})$, which can be calculated by a
closed form expression for the transmission coefficient. ${\cal
L}_n$ is given by

\begin{equation}
{\cal L}_n=\frac{2}{h}\int d\epsilon {\cal
T}_{LR}(\epsilon)(\epsilon-E_F)^n\frac{\partial
f(\epsilon)}{\partial E_F},
\end{equation}
where ${\cal T}_{LR}(\epsilon)$ is the transmission coefficient.
$f(\epsilon)=1/(exp^{(\epsilon-E_F)/k_BT}+1)$. The thermoelectric
coefficients determined by Eq.~(12) are employed to compare with
results of Eqs.~(5)-(7). Because the condition
$\kappa_e/\kappa_{ph} \ll 1$ is readily satisfied in the Coulomb
blockade regime,[9] the optimization of $ZT$ can be
improved by finding the best power factor ($PF=S^2G_e$).

\section{Results and discussion}
\subsection{Effect of destructive QI}
Firstly, we consider TQDM with electron hopping strengths
$t_{LC}=t_{CR}=t_c\neq t_{LR}$. Figure~1 shows the electrical
conductance and Seebeck coefficient as functions of the energy level
of central QD ($\Delta_C=E_C-E_F$) for various coupling strengths,
$t_{LR}$ at $k_BT=1\Gamma_0$. In the small $\Delta_C$ regime
($\Delta_C \le 10\Gamma_0$) $G_e$ is found to be insensitive to
$t_{LR}$, because the charge transport is dominated by the near-resonant tunneling process 
through levels at $E_F+\sqrt{2}t_c$, $E_F$ and $E_F-\sqrt{2}t_c$. On
the other hand, $G_e$ changes significantly with the variation of
$t_{LR}$ in the regime of $\Delta_C/\Gamma_0 \ge 10$. For
$t_{LR}=0.1\Gamma_0$, $G_e$ is vanishingly small at
$\Delta_{C4}=90\Gamma_0$. The minimum of $G_e$ occurs at lower
$\Delta_C$ if $t_{LR}$ is increased. For example, the minimum of
$G_e$ occurs at $\Delta_{C3}=45\Gamma_0$, $\Delta_{C2}=30\Gamma_0$,
and $\Delta_{C1}=22.5\Gamma_0$ for $t_{LR}=0.2\Gamma_0$,
$t_{LR}=0.3\Gamma_0$ and $t_{LR}=0.4\Gamma_0$, respectively. This
behavior is caused by the quantum interference (QI) between two
paths. This direct coupling ($t_{LR}$) between the outer QDs
provides one path.  The long distance coherent tunneling (LDCT)[25]
mediated by the central QD provides another path, which leads to an
effective electron hopping strength $t_{eff}=-t_{LC}t_{CR}/\Delta_C$
between the outer QDs. Such an effective coupling between outer dots
can be used to manipulate the spin entanglement between dots
separated by a long distance.[29-31] We note that the vanishingly
small $G_e$ occurs when the condition of $|t_{eff}|=t_{LR}$ is met.
To depict the behavior of Fig.~1(a), we also calculate $G_e$ by
using the approximate expression, ${\cal T}^1_{LR}(\epsilon)$ for
transmission coefficient as given below.
\begin{eqnarray}
&&{\cal T}^{1}_{LR}(\epsilon) \label{PSB} \\
\nonumber&=&\frac{4\Gamma_L\Gamma_R P_{1}F_{QI}}
{|\mu_1\mu_2\mu_3-t^2_{CR}\mu_1-t^2_{LC}\mu_3-t^2_{LR}\mu_2-2t_{LR}t_{LC}t_{CR}|^2},
\end{eqnarray}
where $\mu_1=\epsilon-E_L+i\Gamma_L$, {$\mu_2=\epsilon-E_C$} and
$\mu_3=\epsilon-E_R+i\Gamma_R$. $P_{1}$ denotes the probability
weight of electron transport through TQDM in an empty state, which
is determined by the one particle occupation number and on-site two
particle correlation functions resulting from electron Coulomb
interactions. {$P_1$ equals to one in the absence of electron
Coulomb interactions.} The numerator
$F_{QI}=(t_{LC}t_{CR}+t_{LR}\mu_2)^2$ causes the QI effect. The
first term and the second term of $F_{QI}$ describe, respectively,
the upper path through the central QD and lower path for direct
hopping between the outer QDs. Due to $t_{LC}=t_{CR}=t_c \neq
t_{LR}$, there are three poles in the denominator of Eq. (13). If we
take $\Gamma=0$, the poles occur at
\begin{eqnarray}
\epsilon_{\pm}
&=&\frac{E_C+E_0+t_{LR}}{2}\pm\frac{1}{2}\sqrt{(E_C-E_0-t_{LR})^2+8t^2_C} \nonumber \\
\epsilon_0&=&E_0-t_{LR}.
\end{eqnarray}
Here, $E_0=E_L=E_R=E_F$ in Eq. (14). For higher symmetry TQDM
($t_{LR}=t_{LC}=t_{CR}=t_c$ and $E_{\ell}=E_0$), we have
$\epsilon_{+}=E_0+2t_c$, $\epsilon_{-}=E_0-t_c$ and
$\epsilon_0=E_0-t_c$.

Once $\Delta_C=E_C-E_F\gg 2t_{c}$, the lowest energy level is given
by $\epsilon_{0}=E_0-t_{LR}$.
Keeping only the resonant channel at $\epsilon_{0}$, we obtain
\begin{equation}
G_e=\frac{2e^2}{h}\frac{\Gamma
\pi}{4k_BT}\frac{(t_{eff}+t_{LR})^2}{(t_{LR}+t_{eff})^2+(\Gamma/2)^2}\frac{P_1}{cosh^2(\frac{t_{LR}}{2k_BT})},
\end{equation}
where $t_{eff}=-t^2_c/\Delta_C$.  This expression can well explain
the destructive QI behavior shown in Fig.~1(a). According to
Eq.~(13), the QI observed in Fig.~1(a) will disappear as $t_{LR}=0$
(the case of SCTQD). In Sanchez et al[32] it was proposed that
destructive QI of two superexchange trajectories can also occur in
SCTQD. Note that such a effect is easily masked by the background
current at finite temperatures. Thus, it is not as robust as the QI
effect described here. Using Eq.~(12) and considering the
$\epsilon_0$ pole of Eq.~(14), we can show that $S=t_{LR}/T$ is
positive and independent on $\Delta_C$. However, $S=t_{LR}/T$ can
not describe the behavior of $S$ shown in Fig.~1(b) (obtained by the
full calculation), which is a function of $\Delta_C$. This implies
that the resonant channels involving $E_C$ can not be ignored for
calculating $S$ in the regime of $t_{eff}/t_{LR}> 1$. The positive
sign of $S$ indicates that the hole diffusion dominates over
electron diffusion at low temperature ($k_BT=1\Gamma_0$). Here holes
are defined as missing electrons in states below $E_F$. The results
of Fig.~1(b) show that the Seebeck coefficient is seriously
suppressed under the destructive QI effect. Such a behavior is quite
different from the general behavior of thermal voltage observed in
QD junction systems.[10-15] (See results of Figs. 5-8 and note that
the minimum of $S$ occurs at the maximum of $G_e$).

To further clarify the above destructive QI behavior, we plot the
electrical conductance and Seebeck coefficient at various
temperatures for $t_{LR}=0.3\Gamma_0$ in Fig.~2. It shows that $G_e$
is suppressed with increasing temperature.  Meanwhile, the
destructive QI effect on $G_e$ is very robust with respect to
temperature variation when $k_BT$ is tuned up to $5\Gamma_0$. The
results of Fig.~2(a) imply that a single-electron QI transistor is
achievable even at room temperature.[33]  The change of sign for $S$
from positive to negative indicates that the resonant channels above
$E_F$ become important with increasing temperature with electron
contribution dominating over hole contribution.  In particular, we
noticed a large enhancement of the negative $S$ peak at
$\Delta_{QI}$ (the value of $\Delta_C$ where $QI$ occurs) as
temperature increases. Namely, the thermal voltage changes sign and
enhances with increasing temperature. We also noticed that peak
value of $|S|$ reaches a maximum near $k_BT=4\Gamma_0$ and decreases
afterwards.

To reveal the importance of electron correlation effects and
understand the interesting behavior of $S$ in Fig.~2(b), we
recalculate the results of Fig.~2 with the procedure of Ref. [9].
There are 32 configurations in the transmission coefficient of Eq.
(12) for electrons with spin $\bar\sigma$ in the electrodes. The
expression of Eq. (13) is for an empty TQDM. The curves in Fig.~3
have one-to-one correspondence to those of Fig.~2. We found
significant differences between Fig.~2(a) and Fig.~3(a), when
$\Delta_C$ is smaller than 20 $\Gamma_0$. This implies that the
electron correlation effects become important when $E_C$ closes to
$E_F$. In particular, two-particle interdot correlation functions.
However, both approaches give similar QI effect with the same value
for $\Delta_{QI}$. {This is because QI is mainly caused by the
single-particle transport process}. Fig.~3(b) also shows enhancement
of $|S|$ at $\Delta_{QI}$ with increasing temperature, but the
degree of enhancement is overestimated compared with the full
calculation. The temperature dependence of $S_{max}$ (at
$\Delta_{QI}$) is plotted in the inset of Fig.~3(b). It is seen that
$S_{max}$ obeys the simple relation $S \approx -\frac{U_{LR}}{T}$
when $k_BT > 2\Gamma_0$.  One can show that the next important
contribution to $S$ is from the resonant channel near
$\epsilon=E_0+U_{LR}-t_{LR}$, but not from $\epsilon_{+}$ given by
Eq.~(14). Note that the contribution of this resonant channel to
$G_e$ is small compared to the $\epsilon_0$ channel. To demonstrate
the effect of electron Coulomb interactions, we plot the curve with
red triangle marks for $U_{\ell}=U_{\ell,j}=0$ at $k_BT=2\Gamma_0$.
Now, $S_{max}=t_{LR}/T$ becomes very small in the absence of
electron Coulomb interactions. This proves that the enhancement of
$|S_{max}|$ results from the resonant channels involving electron
Coulomb interactions. This also explains the enhancement of
$|S_{max}|$ shown in Fig.~2(b).

According to the results of Figs.~(1) and (2), the destructive QI
effect can blockade the charge transport. Next, we clarify how QI
influences electron heat transport. Fig.~4 shows the electron
thermal conductance ($\kappa_e$) and Lorenz number
$(L_z=\kappa_e/(G_eT))$ as functions of $\Delta_C$. The physical
parameters adopted in the calculation are the same as those for
Fig.~2. Like $G_e$, the electron heat transport can be modulated by
tuning the central QD energy level. We see that the electron thermal
conductance is also vanishingly small at $\Delta_{QI}=30\Gamma_0$
due to the destructive QI. There is the Wiedemann-Franz law in
metals to link two physical quantities ($G_e$ and $\kappa_e$). Here,
we recheck the Wiedemann-Franz law by calculating the Lorenz number
in Fig.~4(b). A highly temperature-dependent behavior of $L_z$
occurs at $\Delta_{QI}$. The fact that $L_z$ larger than one
indicates that it is easier to block charge transport via
destructive QI than the electron heat transport. In metals the
Lorenz number is a universal constant
$L_z=\frac{\pi^2}{3}(k_B/e)^2$, which is independent of T. Our
results obviously violet the Wiedemann-Franz law near
$\Delta_C=\Delta_{QI}$ for all temperatures. The violation of
Wiedemann-Franz law is a typical phenomenon for nanostructures with
discrete energy levels.[3] The QI effects on the TE properties are
illustrated by using QDMs with N=3 as an example. The asymmetrical
features of $t_{LC}=t_{CR}\neq t_{LR}$, $U_{LC}=U_{CR}\neq U_{LR}$
and $E_L=E_R\neq E_C$ considered in TQDM can be used to reveal the
effects of of QD size fluctuation and nonuniform interdot separation
 on the charge transport of QDSL system.[6,8]

\subsection{Orbital filling effects}
Many theoretical studies have used the first-principles method
(within mean-field approximation) to predict the TE coefficient of
realistic molecules with orbital below the $E_F$ of
electrodes.[22-25] Such a mean-field approach does not take full
account of the electron correlation effects. Therefore, the
predicted $G_e$, S and PF can not reveal the realistic TE properties
of molecule junction systems in the Coulomb blockade regime. To
reveal the orbital charge-filling (orbital below $E_F$) effects on
the thermoelectric properties of TQDM, we plot the total occupation
number ($N=\sum_{\sigma}(N_{L,\sigma}+N_{c,\sigma}+N_{R,\sigma})$),
$G_e$ and S as functions of gate voltage $V_g$
($E_{\ell}=E_F+30\Gamma_0-eV_g$) at $\Gamma=0.3\Gamma_0$,
$t_{\ell,j}=3\Gamma_0$ and $U_{\ell,j}=30\Gamma_0$ for various
temperatures in Fig.~5. Here, the QD energy levels are placed at
$30\Gamma_0$ above $E_F$ when $eV_g=0$. The average total occupation
number ($N$) display an staircase behavior with six plateaus. The
step edges are broadened as temperature increases. The height of
each staircase is equal to one. The onsets of these plateaus are
roughly determined by $U_{\ell,j}$ and $U_{\ell}$.  Therefore, the
electron number of TQDM is tuned from one to six with increasing
gate voltage. The spectrum of $G_e$ exhibits a Coulomb oscillation
behavior with respect to gate voltage. Six main peaks of $G_e$
labeled by $\epsilon_n$ correspond to the particle-addition energy
of TQDM with different many-body states. For $k_BT=1\Gamma_0$, there
are several insulating states (with vanishingly small $G_e$). Such
insulating states are not due to the QI effect, but the Coulomb
blockade effect. The regime of vanishingly small $G_e$ is related to
a very small transmission coefficient resulting from the absence of
resonant levels near $E_F$. The insulating state of a single benzene
molecule calculated by the DFT method was suggested from the
destructive QI effect.[22-24] The suggestions from Refs. [22-24] are
unclear due to the deficiency of the mean-field theory (DFT).
Although the behavior of $G_e$ with respect to $k_BT$ is a typical
behavior that the magnitude of $G_e$ peak is suppressed with
increasing temperature, meanwhile the width of each peak becomes
broadened, we find a more dramatic reduction of $\epsilon_6$ peak
when $k_BT$ is tuned from $1\Gamma_0$ to $2\Gamma_0$. In addition,
the peak position is shifted with increasing temperature. Several
secondary peaks resulting from excite states around these main peaks
are washed out by increasing temperature. The separation between
$\epsilon_3$ and $\epsilon_4$ is the so called "Coulomb gap" (these
two peaks are separated by the intradot Coulomb interaction
$U_{\ell}$).[25] Using the middle Coulomb gap as a reference point,
the $G_e$ spectrum does not maintain the mirror symmetry. This is
different from that of DQDs.

The Seebeck coefficient shown in Fig.~5(c) also shows an oscillatory
behaviors with respect to gate voltage. When QD energy levels are
above $E_F$ the Seebeck coefficients are negative, which indicates
that electrons dominate the diffusion process of thermoelectric
properties. When QD energy levels near $E_F$, $S$ almost vanishes
due to the electron-hole balance. The positive $S$ indicates that
holes through resonant level below $E_F$ become majority carriers.
Such bipolar effects are the interplay between holes and electrons.
When the resonant levels corresponding to the two-electron and
three-electron states of TQDM ($\epsilon_2$ and $\epsilon_3$) are
near $E_F$, once again the Seebeck coefficients become very small.
We note that when $G_e$ reaches the maximum, the Seebeck coefficient
($S=\Delta V/\Delta T$) reaches zero. This is different from that of
Fig.~2. The Seebeck coefficient near the onsets of plateaus for
$N=1, 3, 4$, and 6 are highly enhanced with increasing temperature.
Like the $G_e$ spectrum, the Seebeck coefficient does not exhibit a
mirror symmetry. We find prominent secondary oscillatory structures
of $S$ (clearly noticeable at $k_BT=1\Gamma_0$), which arise from
the secondary resonant levels around the main peaks of $G_e$.
Compared to the spectrum of $G_e$ at low temperature, the spectrum
of $S$ can clearly resolve the structures arising from the excited
states (secondary resonances) of TQDM. Consequently, the measurement
of $S$ is a powerful means to reveal various configurations of TQDM
arising from the many-body effect. Unlike the case of metallic
QDs,[34] the spectrum of $S$ does not show the periodically
oscillating oscillatory structure with respect to the gate voltage
due to $U_{\ell}\neq U_{\ell,j}$ and electron correlation effects.
The interdot Coulomb interactions as well as intradot Coulomb
interactions play significant roles for electron transport.

To examine the interdot Coulomb interaction effects, Figure~6 shows
(a) the electrical conductance, (b) Seebeck coefficient and (c)
power factor as functions of $V_g$ for three different values of
$U_{\ell,j}$ at $k_BT=1\Gamma_0$. Other physical parameters are the
same as those of Fig.~5. From the results of $G_e$, we find that the
interdot Coulomb interactions not only influence the peak positions,
but also change the magnitude of each peak. For $U_{\ell,j}=0$, the
$G_e$ spectrum shows only four peaks (see blue dashed line). The
first two peaks (separated by 3$t_c$) correspond to the resonant
levels $\epsilon_{BD}=E_0-t_c$ (bonding state) and
$\epsilon_{AB}=E_0+2t_c$ (antibonding state). It is noted that the
first $G_e$ peak at $\epsilon_{BD}$ is significantly reduced and the
second peak at $\epsilon_{AB}$ is totally suppressed when
$U_{\ell,j}$ becomes finite.

The peak structures of the $G_e$ spectrum for
$U_{\ell,j}=15\Gamma_0$ is quite similar to that for
$U_{\ell,j}=30\Gamma_0$, except the spacings between peaks are
different, which leads to very different behavior in the Seebeck
coefficient. Previous theoretical works predicted the thermal power
(or Seebeck coefficient) spectra without considering interdot
Coulomb interactions.[15]. This is inadequate for studying
semiconductor QD molecules with nanoscale separation between QDs.
Our calculations indicate that to achieve large maximum power factor
(PF$_{max}$) in TQDM with homogenous QD energy levels and electron
hopping strengths, the full-filling condition (with six-electron
state) is preferred. This is due to the fact that three resonant
channels associated states localized at dot L, C, and R (with
energies at $E_L+U_0+2U_{LC}+2U_{LR}$, $E_C+U_0+2U_{LC}+2U_{CR}$,
and $E_R+U_0+2U_{LR}+2U_{CR}$, respectively) are all aligned. It is
interesting to see what would happen in other QDM structures when
the resonant conditions are changed.

Figure~7  shows the electrical conductance ($G_e$), Seebeck
coefficient ($S$) and power factor (PF) of DQD as functions of $V_g$
at different temperatures. Physical parameters are the same as those
for Fig.~5. For DQD, the  four $G_e$ peaks correspond to
$\epsilon_1=E_0-t_{LR}$,
$\epsilon_2=E_0+t_{LR}+\frac{U_0+U_{LR}}{2}-\frac{1}{2}\sqrt{(U_0-U_{LR})^2+16t^2_{LR}}$,
$\epsilon_3=E_0-t_{LR}+\frac{U_0+3U_{LR}}{2}+\frac{1}{2}\sqrt{(U_0-U_{LR})^2+16t^2_{LR}}$
and $\epsilon_4=E_0+U_0+2U_{LR}+t_{LR}$. The $G_e$ spectrum shows a
mirror-symmetry behavior with respect to the middle of the Coulomb
gap. Due to the symmetry of DQD structure, the PF spectrum also
shows the mirror symmetry. The maximum PF occurs at either
orbital-depletion ($N \le 1$) or orbital-filling ($N=4$) condition
for DQD. Note that unlike $G_e$ and PF, $S$ does not show the mirror
symmetry.}

Figure~8  shows $G_e$, $S$ and PF of SCTQD as functions of $V_g$ at
different $T$. For simplicity,  we adopt $t_{LR}=0$ and $U_{LR}=0$,
since they are typically much smaller than the corresponding
parameters related to the central dot. The $G_e$ spectrum in Fig.~8
does not show the mirror symmetry as a result of the inhomogeneous
interdot Coulomb interactions ($U_{LC}=U_{CR}\neq U_{LR}$). The
peaks labeled $\epsilon_N$ ($N=1,\cdots,6$) correspond to the
transitions where the total particle numbers in the SCTQD changes
from $N-1$ to $N$ with $\epsilon_1$, $\epsilon_2$, and $\epsilon_4$
being more prominent. The $\epsilon_1$ peak comes from electrons
tunneling through the empty SCTQD via the resonant level
$\epsilon_1=E_0-\sqrt{2}t_c$. The $\epsilon_2$ peak (
$\epsilon_2=E_0+\sqrt{2}t_c+U_{LR}$) corresponds to charge transport
in the presence of another electron localized in one of the two
outer dots. (Here we have chosen $U_{LR}=0$) This is verified by
examining $N_{C,\sigma}$, which is negligible until $eV_g$ is near
$\epsilon_3$. $\epsilon_4=E_0+U_0+U_{LC}(U_{CR})$ corresponds to the
injection of an electron into the left (right) dot of SCTQD with all
three dots each filled with one electrons, while
$\epsilon_5=E_0+U_0+U_{LC}(U_{CR})+2t_{eff}$ corresponds to the
addition of another electron in either left or right dot to the
configuration associated with $\epsilon_4$, where
$t_{eff}=t_{LC}t_{CR}/(E_F-(E_C+2U_{LC}+2U_{CR}))=0.9\Gamma_0$ is
the effective hopping strength mediated by the center dot.
Therefore, the splitting between $\epsilon_4$ and $\epsilon_5$ can
be observed by increasing $t_{\ell,j}$. The Seebeck coefficient
spectrum can reveal fine structures not resolved in the $G_e$
spectrum. For example, the $\epsilon_3$ and $\epsilon_6$  peaks are
hardly noticeable in $G_e$ but clearly observable in $S$. The peak
at $\epsilon_3=E_0+U_{LC}+U_{CR}$ is caused by the charging of the
center dot with two outer dots each filled with one electron and
$\epsilon_6$ describes the charging of SCTQD with a second electron
into the center QD with two outer dots each filled with two
electrons. In strong contrast to TQDM, both $G_e$ and  PF of SCTQD
are very small at the maximum filling condition ($N=6$), which
occurs at $eV_g=250\Gamma_0$ where $E_C+U_0+2U_{LC}+2U_{CR}$ is
aligned with $E_F$. This is because the three resonant channels
associated states localized at dot L, C, and R (with energies at
$E_L+U_0+2U_{LC}+2U_{LR}$, $E_C+U_0+2U_{LC}+2U_{CR}$, and
$E_R+U_0+2U_{LR}+2U_{CR}$, respectively) are  misaligned, since
$U_{LR}=0$ (or $U_{LR}\ll U_{CR}$ in general). Consequently, charge
transport is seriously suppressed for QD molecule with
orbital-filling condition, which is expected to be a general feature
in serially coupled QDMs with more dots. For the present SCTQD, the
maximum PF occurs at the combined $\epsilon_4$ and $\epsilon_5$ peak
as shown in Fig.~8. In this case, the charge transport is enabled
via the long distance coherent tunneling mechanism, which means that
the localized electron of the left dot can be transferred to the
right dot via an effective hopping strength mediated by the central
dot.[25] When we consider $U_{LR}$ or tune $t_{\ell,j}$ away from
$3\Gamma_0$ (not shown here), the maximum PF for SCTQD would occur
at the orbital-depletion condition.

\section{Summary}
To reveal the QI and orbital filling effects o the charge/heat
transport in QDSL and molecule junction system, we have
theoretically investigated the TE coefficients of TQDM with a full
many-body solution. Our theoretical work can serve as a guideline
for the design of nanoscale TE devices. The destructive QI of $G_e$
is found to be very robust with respect to temperature variation.
This implies that it is possible to achieve the destructive QI of
$G_e$ at high temperatures. Furthermore, the Seebeck coefficient can
be made large with a vanishingly small electrical conductance when
temperature increases. From the results of various QDMs with a few
electrons, we have demonstrated that the measurement of Seebeck
coefficient, $S$ is a powerful tool to reveal the many-body effect
of nanostructures, since $S$ is influenced much more than $G_e$.To
achieve maximum power factor in the case of TQDM  the
orbital-filling situation is preferred (with $N$=6). However, for
SCTQD in general it is preferable to the orbital-depletion situation
with ($N \le 1$) and likely for serially coupled QDMs with more
dots. It is desirable to compare current results including all Green
functions and electron correlations with those calculated by the
Hartree-Fock approximation (HFA) [22-24] to examine the validity of
mean-field approach. Such a comparison for SCTQD has been carried
out.[35] We found that HFA works well only for the orbital-depletion
situation and high temperatures.



\begin{flushleft}
{\bf Acknowledgments}\\
\end{flushleft}
This work was supported by the National Science Council of the
Republic of China under Contract Nos. NSC 103-2112-M-008-009-MY3 and
NSC 101-2112-M-001-024-MY3.

\mbox{}\\
E-mail address: mtkuo@ee.ncu.edu.tw\\
E-mail address: yiachang@gate.sinica.edu.tw\\

\setcounter{section}{0}

\renewcommand{\theequation}{\mbox{A.\arabic{equation}}} 
\setcounter{equation}{0} 

\mbox{}\\

\newpage
\clearpage

\begin{figure}[h]
\centering
\includegraphics[angle=-90,scale=0.3]{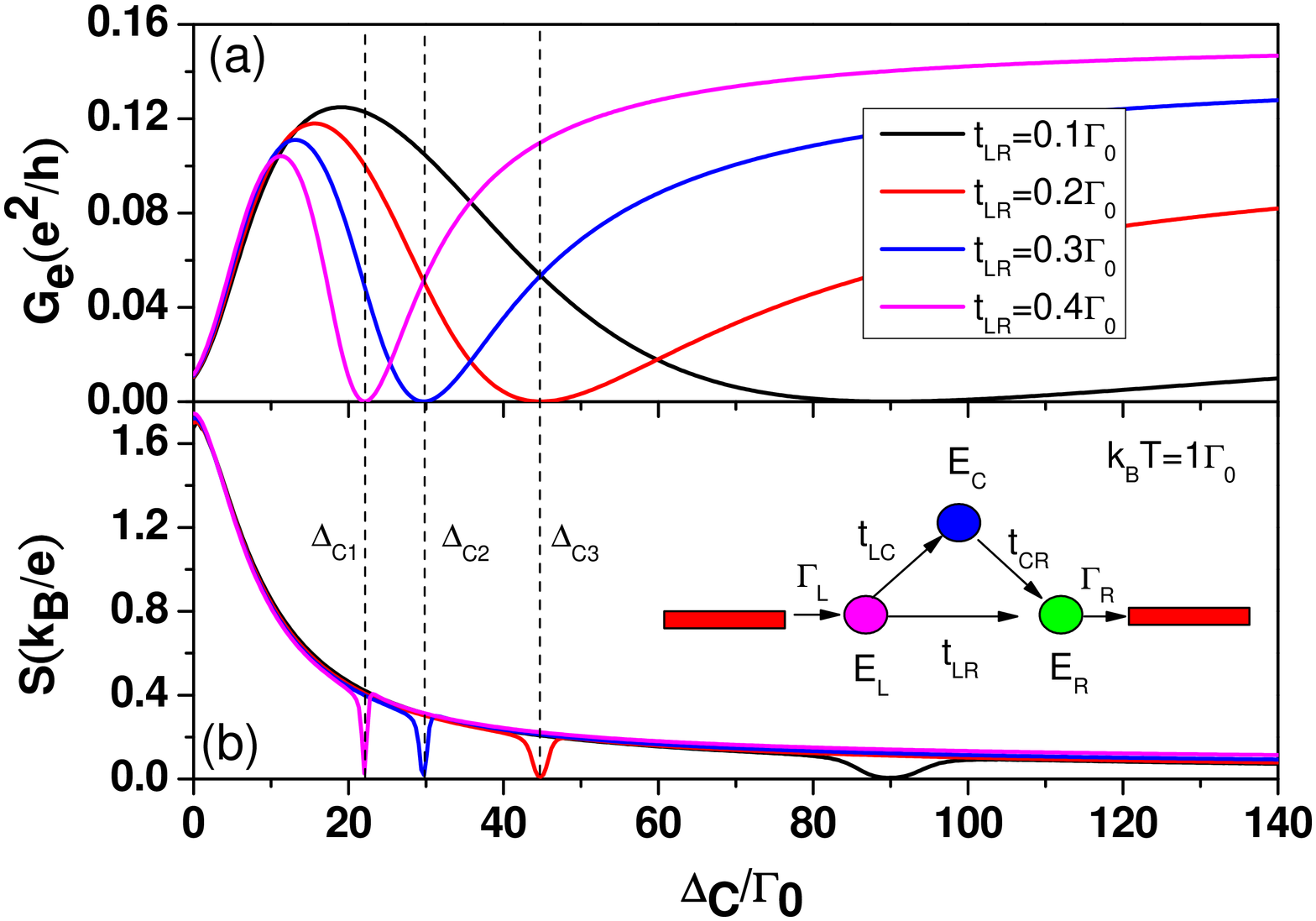}
\caption{(a) Electrical conductance ($G_e$) and (b) Seebeck
coefficient (S) of TQDM as a function of central QD energy
($\Delta_C=E_C-E_F$) for different $t_{LR}$ strengths at
$E_L=E_R=E_F$, $t_{LC}=t_{CR}=t_c=3\Gamma_0$ and $k_BT=1\Gamma_0$.
We assume $U_{LC}=U_{CR}=30\Gamma_0$, $U_{LR}=10\Gamma_0$,
$U_{\ell}=U_0=100\Gamma_0$ and
$\Gamma_L=\Gamma_R=\Gamma=0.3\Gamma_0$.}
\end{figure}

\begin{figure}[h]
\centering
\includegraphics[angle=-90,scale=0.3]{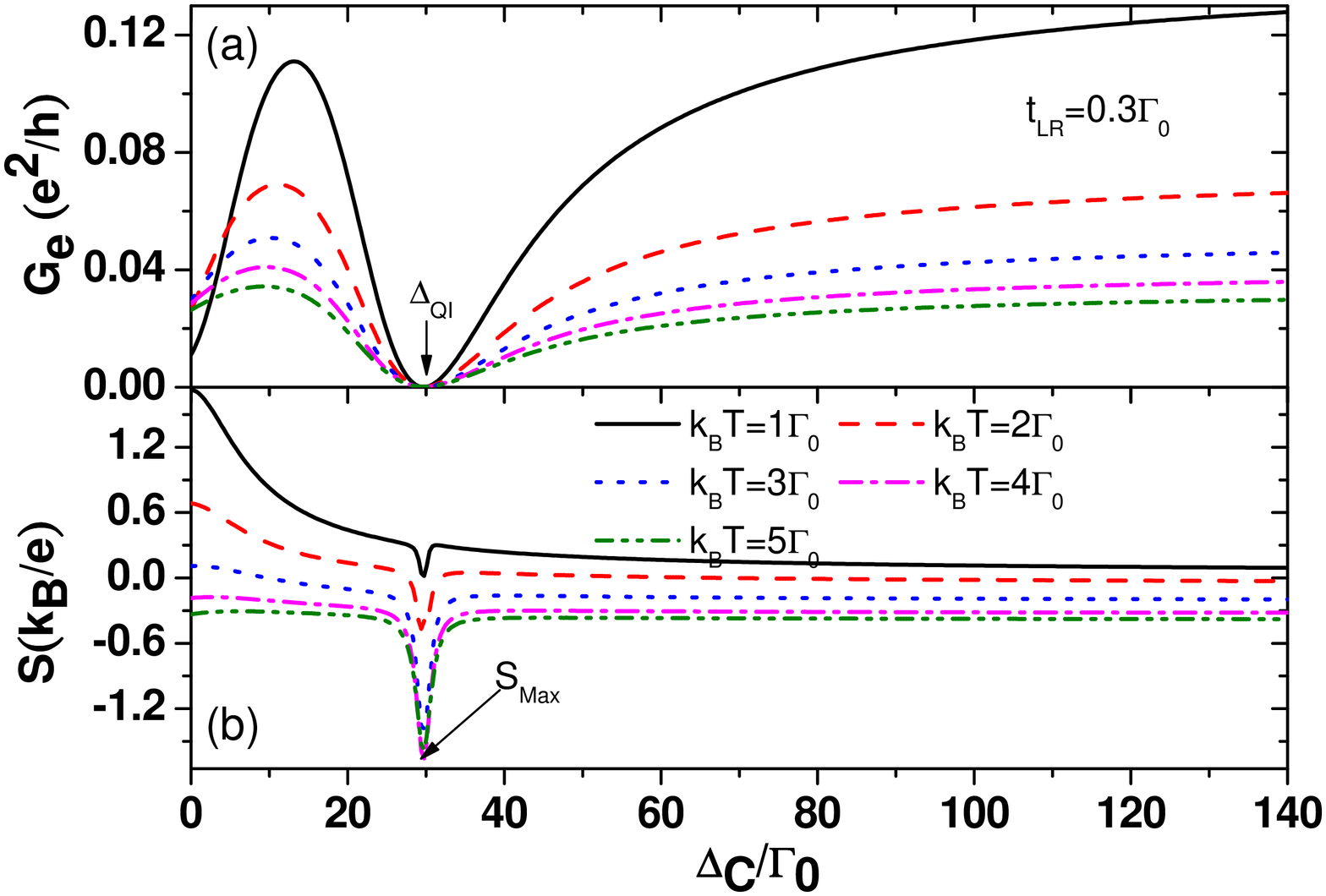}
\caption{(a) Electrical conductance ($G_e$) and (b) Seebeck
coefficient (S) of TQDM as a function of central QD energy
($\Delta_C=E_C-E_F$) at
$t_{LR}=0.3\Gamma_0$ for different temperatures. Other physical parameters are the same as
those of Fig. 1.}
\end{figure}

\begin{figure}[h]
\centering
\includegraphics[angle=-90,scale=0.3]{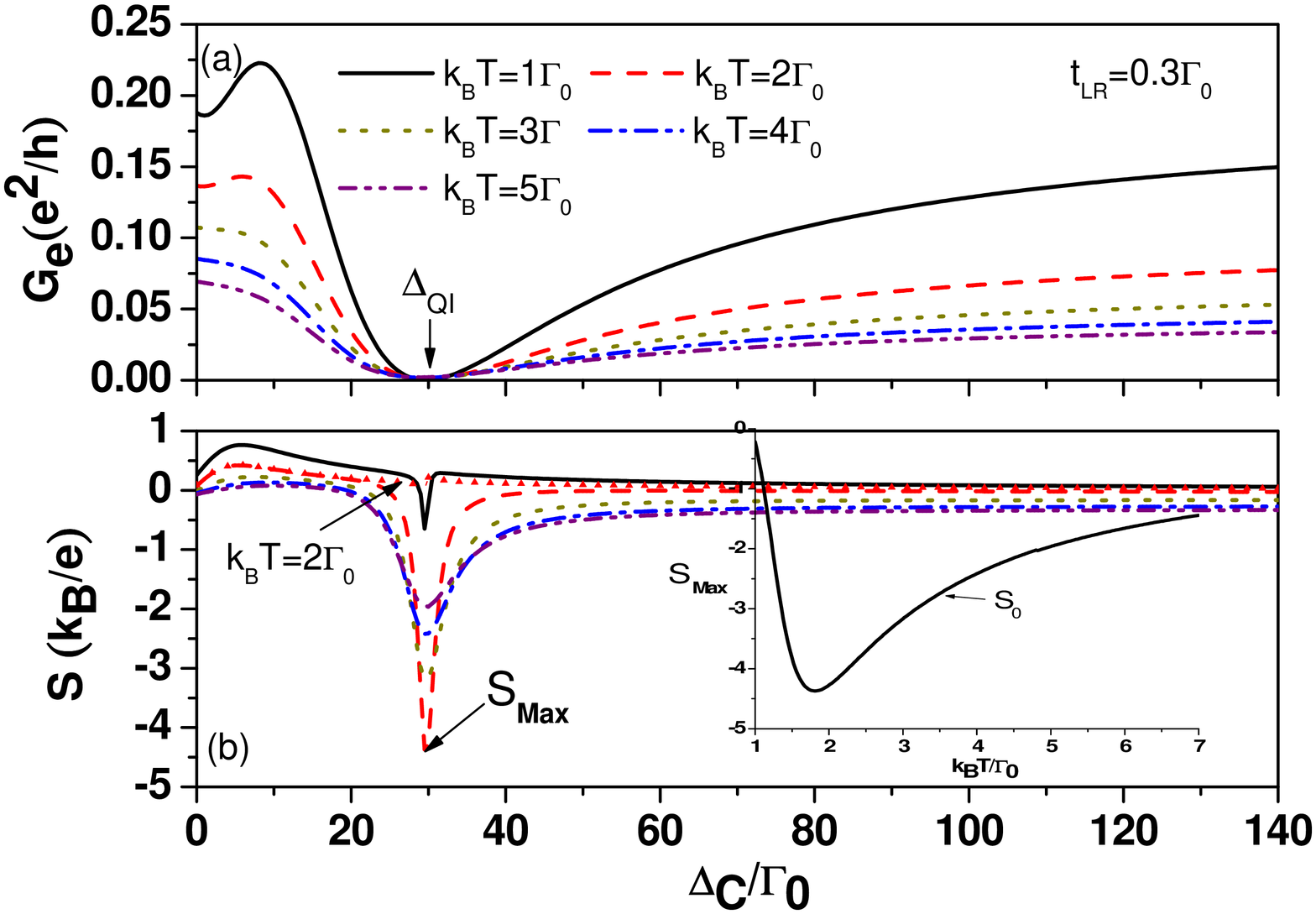}
\caption{The curves of Fig.~3 are one to one corresponding to those
of Fig. 2. The calculation of Fig. 3 only considers the single
particle occupation numbers and on-site two particle correlation
functions.}
\end{figure}

\begin{figure}[h]
\centering
\includegraphics[angle=-90,scale=0.3]{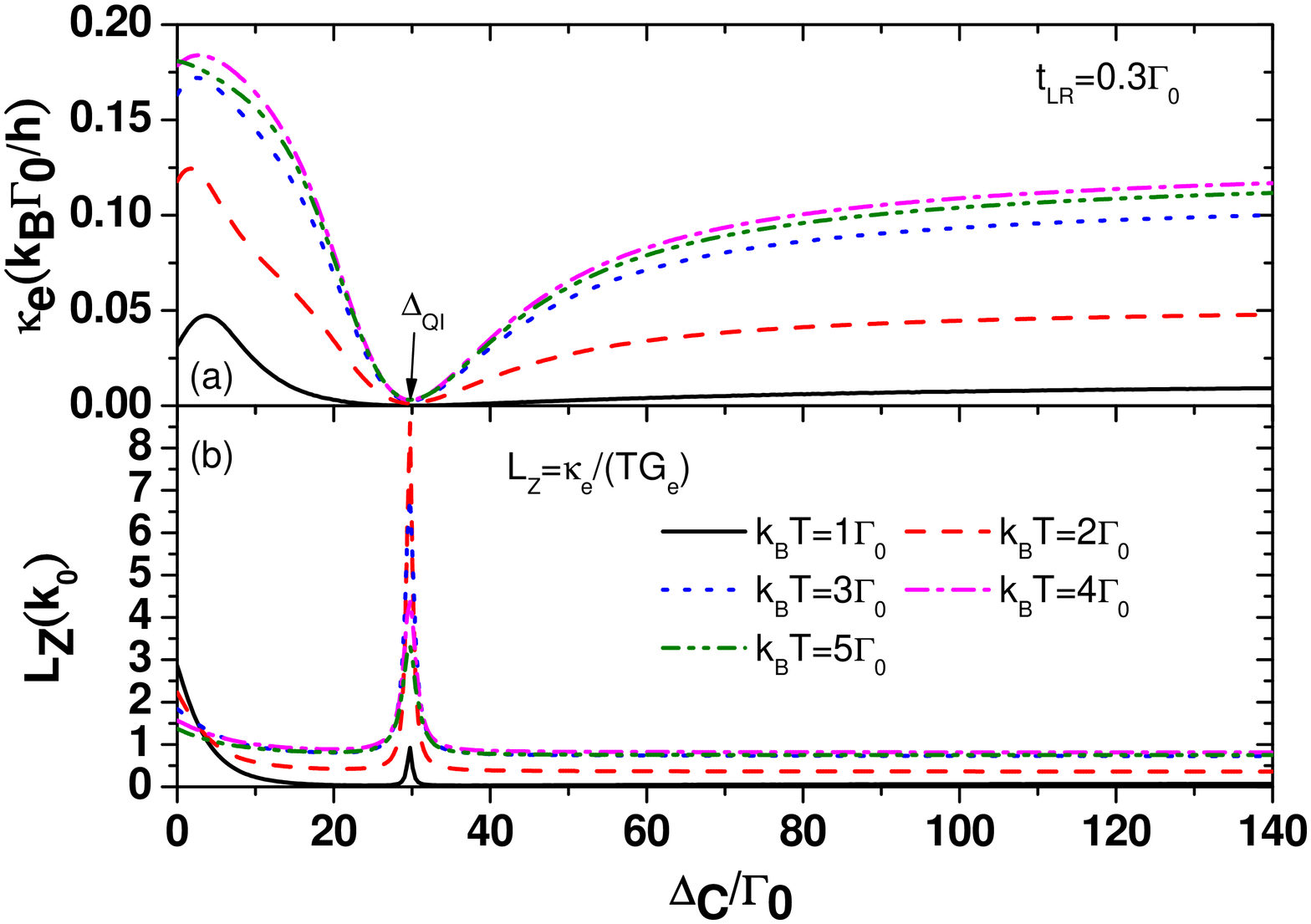}
\caption{(a) Electron thermal conductance ($\kappa_e$), and (b)
Lorenz number($\kappa_e/(G_eT)$) as a function of central QD energy
for different temperatures. The curves of Fig. 4 correspond to those
of Fig. 2. $k_0=k_B^2/e^2$.}
\end{figure}

\begin{figure}[h]
\centering
\includegraphics[angle=-90,scale=0.3]{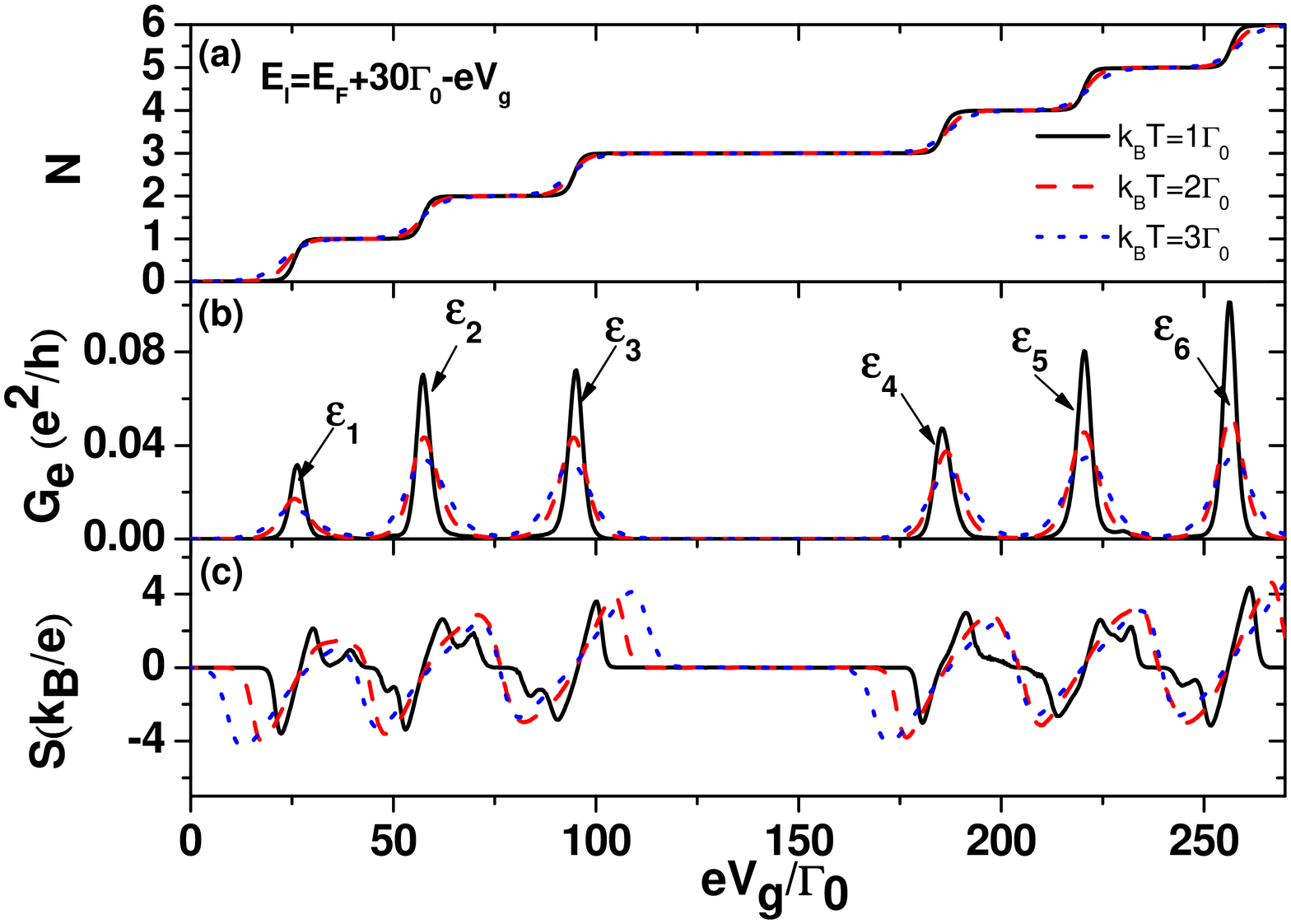}
\caption{(a) Total occupation number, (b) electrical conductance
($G_e$), and (c) Seebeck coefficient (S) as a function of QD energy
($E_{\ell}=E_F+30\Gamma_0-eV_g$) at
$t_{\ell,j}=3\Gamma_0$ and $U_{\ell,j}=30\Gamma_0$  for different temperatures. Other physical
parameters are the same as those of Fig. 1.}
\end{figure}

\begin{figure}[h]
\centering
\includegraphics[angle=-90,scale=0.3]{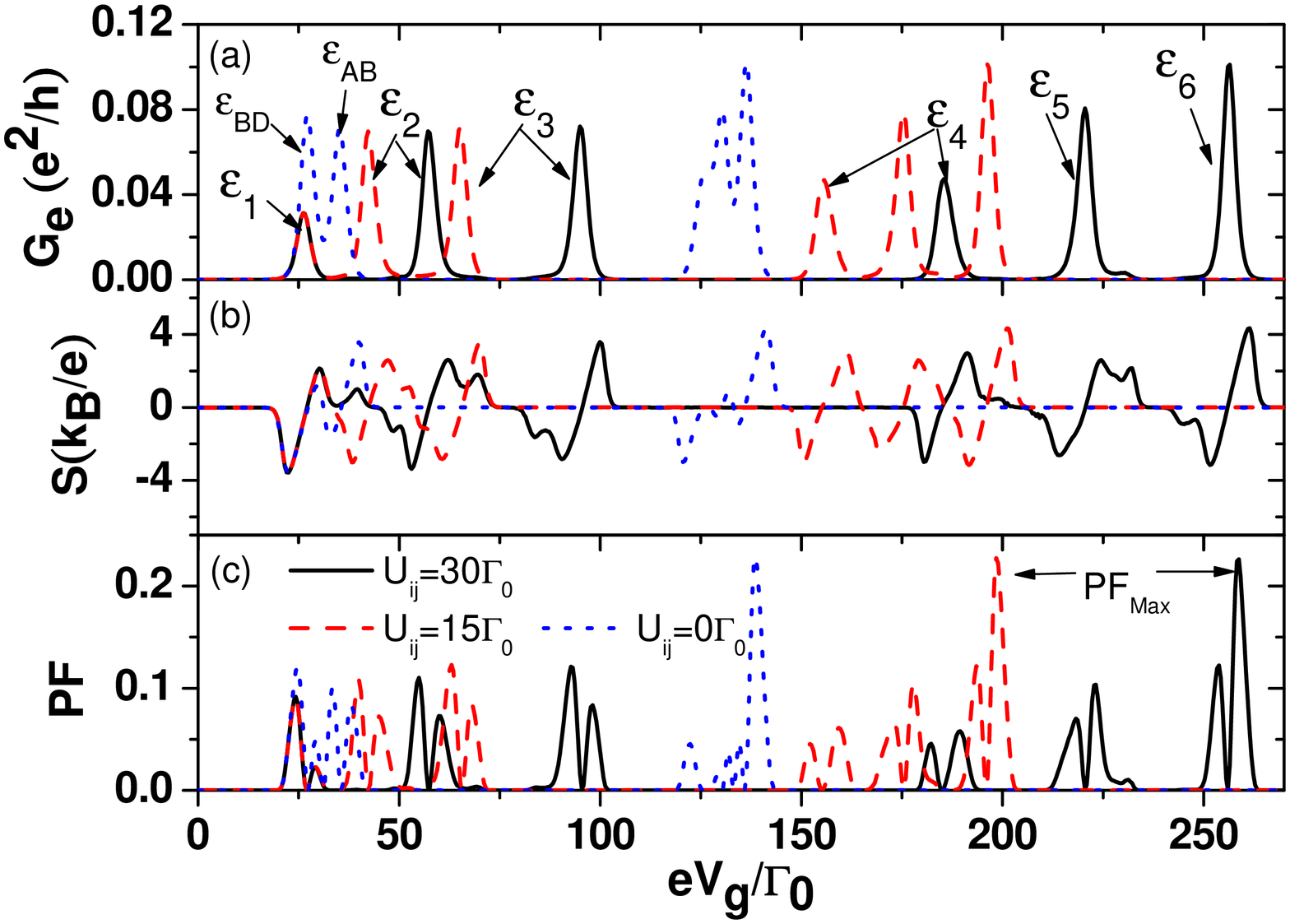}
\caption{(a) Electrical conductance ($G_e$), (b) Seebeck coefficient
(S) and (c) power factor as a function of QD energy
($E_{\ell}=E_F+30\Gamma_0-eV_g$) at $t_{\ell,j}=t_c=3\Gamma_0$ and
$k_BT=1\Gamma_0$ for different interdot Coulomb
interactions. Other physical parameters are the same as those of
Fig. 5.}
\end{figure}

\begin{figure}[h]
\centering
\includegraphics[angle=-90,scale=0.3]{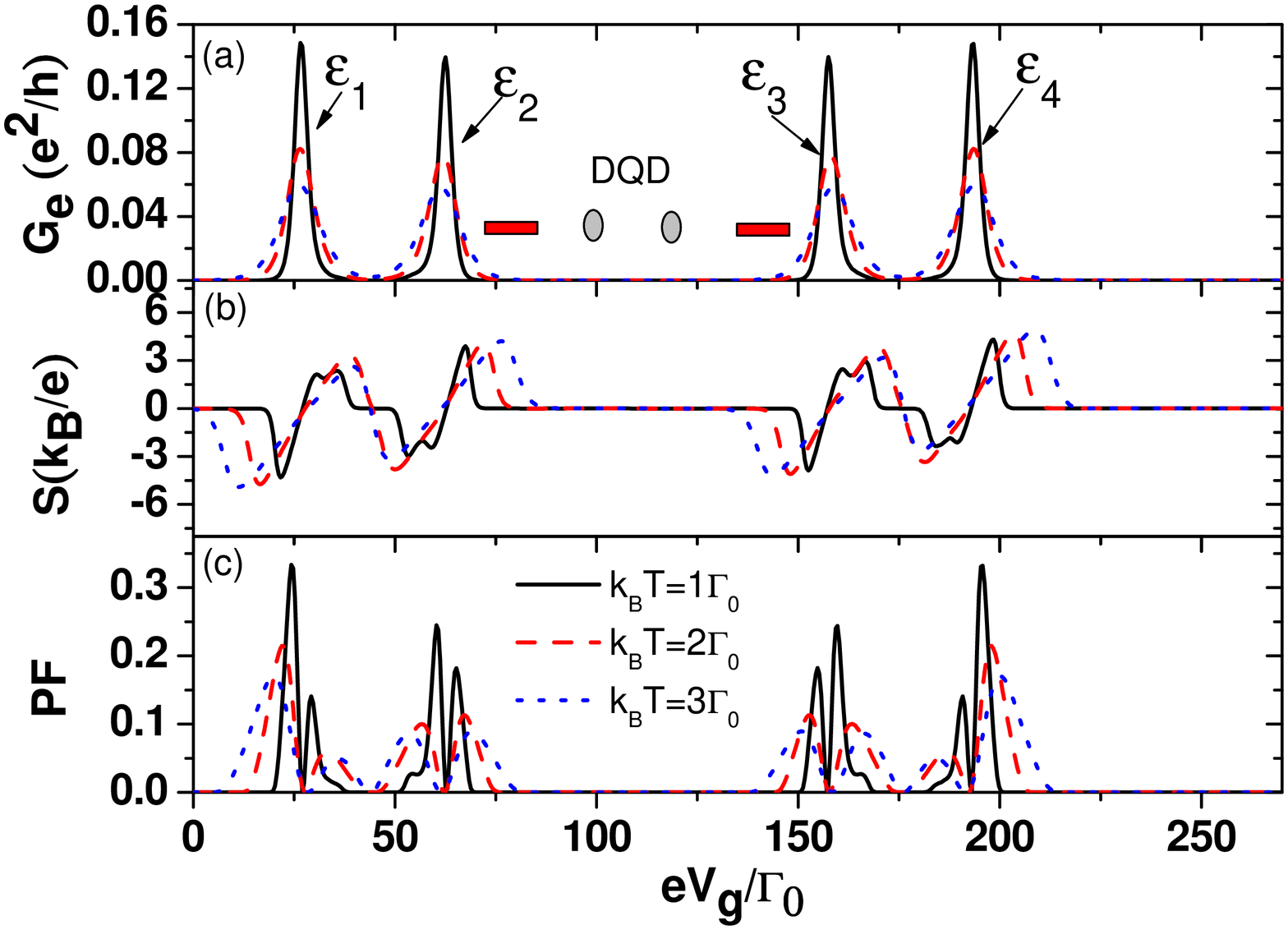}
\caption{Electrical conductance ($G_e$), Seebeck coefficient (S) and
power factor of DQD as functions of QD energy
($E_{\ell}=E_0=E_F+30\Gamma_0-eV_g$) for different temperatures.
Other physical parameters are the same as those of Fig. 5.}
\end{figure}

\begin{figure}[h]
\centering
\includegraphics[angle=-90,scale=0.3]{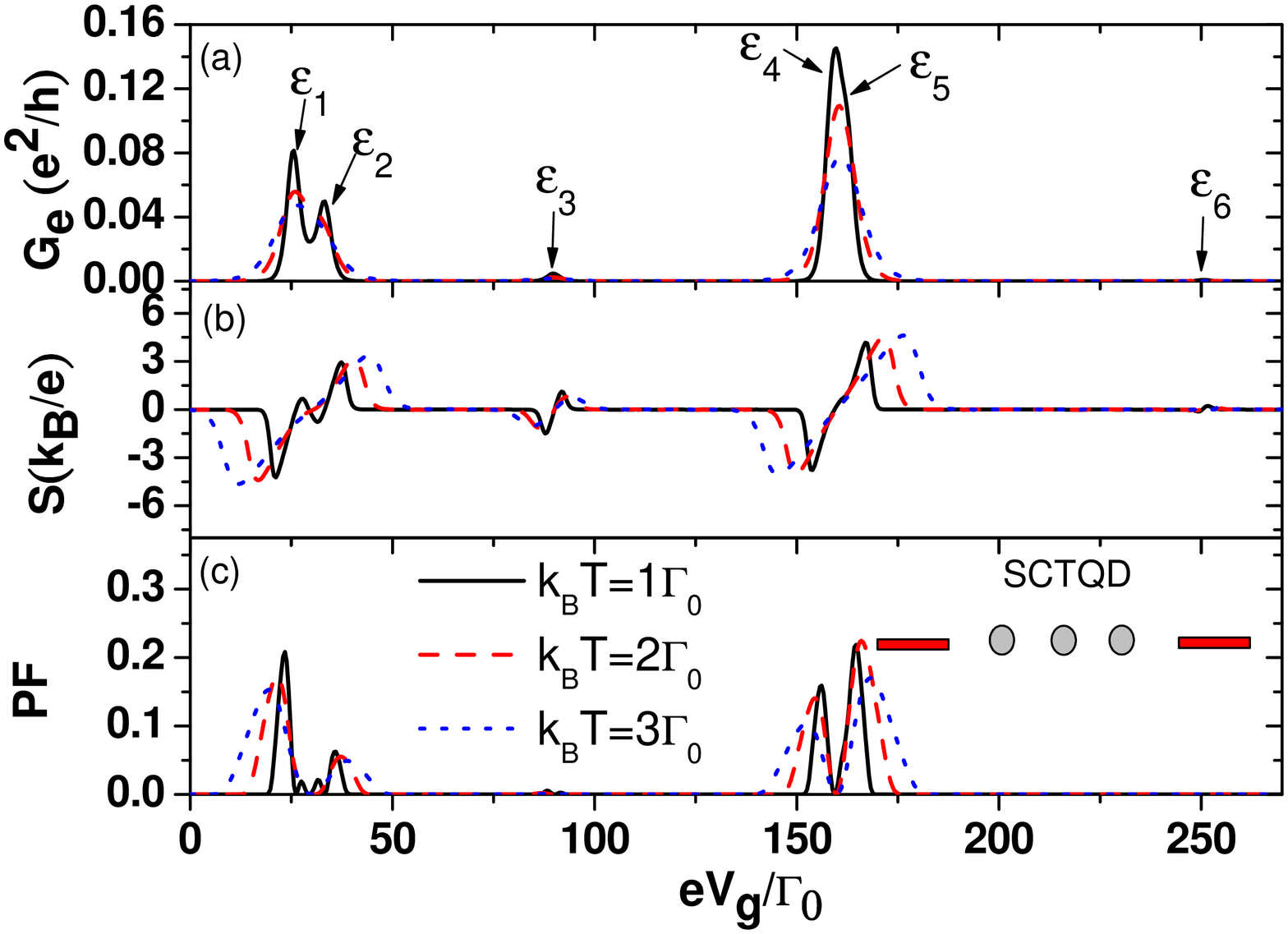}
\caption{Electrical conductance ($G_e$), Seebeck coefficient (S) and
power factor of SCTQD as functions of QD energy
($E_{\ell}=E_F+30\Gamma_0-eV_g$) for different temperatures.
$U_{LR}=0$ and $t_{LR}=0$. Other physical parameters are the same as
those of Fig. 5. }
\end{figure}

\clearpage


\begin{thebibliography}{50}

\bibitem[1]{Min} A. J. Minnich, M. S. Dresselhaus, Z. F. Ren,
and G. Chen, Energy Environ Sci \textbf{2} 466 (2009).

\bibitem[2]{Zeb} M. Zebarjadi, K. Esfarjania, M.S. Dresselhaus, Z.F. Ren, and G.
Chen, Energy Environ Sci \textbf{5} 5147 (2012).


\bibitem[3]{Mah} G. Mahan, B. Sales, and J. Sharp, Physics Today
\textbf{50} (3) 42 (1997).

\bibitem[4]{Ven} R. Venkatasubramanian, E. Siivola, T. Colpitts, and B. O'Quinn, Nature  \textbf{413}
597 (2001).

\bibitem[5]{Bou} A. I. Boukai, Y. Bunimovich, J. Tahir-Kheli, J. K.
Yu, W. A. Goddard III, and J. R. Heath, Nature \textbf{451} 168
(2008).

\bibitem[6]{Har} T. C. Harman, P. J. Taylor, M. P. Walsh, and B. E.
LaForge, Science \textbf{297} 2229 (2002).


\bibitem[7]{Hsu} K. F. Hsu, S. Loo, F. Guo,W. Chen, J. S. Dyck, C. Uher, T. Hogan,
E. K. Polychroniadis, and M. G. Kanatzidis, Science \textbf{303} 818
(2004).

\bibitem[8]{Dre} M. S. Dresselhaus, G. Chen, M. Y. Tang, R. Yang, H.
Lee, D. Wang, Z. Ren, J. P. Fleurial, and P. Gogna, Adv.
Mater.\textbf{ 19},  104 (2007).

\bibitem[9]{Kuo3} D. M. T. Kuo and Y. C. Chang, Nanotechnology \textbf{24}, 175403  (2013).

\bibitem[10]{Mur} P. Murphy, S. Mukerjee, and J. Moore, Phys. Rev. B
\textbf{78}, 161406 (2008).

\bibitem[11]{Kuo2} David M T Kuo and Y. C. Chang, Phys. Rev. B \textbf{81},
205321 (2010).

\bibitem[12]{Liu} J. Liu,  Q. F. Sun, and X. C. Xie, Phys. Rev. B \textbf{81},
245323 (2010).

\bibitem[13]{Wie} M. Wierzbicki, and R. Swirkowicz, Phys. Rev. B \textbf{84},
075410 (2011).

\bibitem[14]{San} D. Sanchez, and L. Serra, Phys. Rev. B \textbf{84},  201307
(2011).

\bibitem[15]{Tro} P. Trocha, and J. Barnas, Phys. Rev. B \textbf{85},  085408 (2012).



\bibitem[16]{Chen1} G. Chen, and C. L. Tien, J. Thermophys.
Heat Tansfer \textbf{7},  311 (1993).

\bibitem[17]{Chen6} G. Chen, J. Heat. Transfer 119, 220 (1997).

\bibitem[18]{Li} D. Y. Li, Y. Y. Wu, P. Kim, L. Shi, P. D. Yang, and A. Majumdar, Appl. Phys. Lett.
\textbf{83}  2934 (2003).

\bibitem[19]{Hoc} A. I. Hochbaum, R. K. Chen, R. D. Delgado, W. J. Liang, E. C. Garnett, M. Najarian,
A. Majumdar, and P. Yang, Nature \textbf{451},  163-U5 (2008).

\bibitem[20]{Ren} R. Chen, A. I. Hochbaum, P. Murphy, J. Moore, P.
D. Yang, and A. Majumdar, Phys. Rev. Lett. \textbf{101}, 105501
(2008).

\bibitem[21]{Nik} D. L. Nika, E. P. Pokatilov, A. A. Balandin, V. M. Fomin, A.
Rastelli, and O. G. Schmidt, Phys. Rev. B \textbf{84}, 165415
(2011).

\bibitem[22]{Ber} J. P. Bergfield, and C. A. Stafford, Phys. Rev. B
\textbf{79}, 245125 (2009).

\bibitem[23]{Ber1} J. P. Bergfield and C. A. Stafford, Nano Letters
\textbf{9}, 3072 (2009).

\bibitem[24]{Ber2} J. P. Bergfield, M. A. Solis, and C. A.
Staffford, ACS Nano \textbf{4}, 5314 (2010).

\bibitem[25]{Che} C. C. Chen, Y. C. Chang and David M T Kuo,
Phys. Chem. Chem. Phys. \textbf{17}, 6606 (2015).

\bibitem[26]{Hau} H. Haug and A. P. Jauho, Quantum Kinetics in Transport and Optics
of Semiconductors (Springer, Heidelberg, 1996).

\bibitem[27]{Jau} A. P. Jauho, N. S. Wingreen and Y. Meir, Phys.
Rev.  B \textbf{50}, 5528 (1994), and references therein.

\bibitem[28]{Ogu} A. Oguri, S. Amaha, Y. Nishikawa, T. Numata, M.
Shimamoto, A. C. Hewson and S. Tarucha, Phys. Rev. B \textbf{83},
205304 (2011).

\bibitem[29]{Bus} M. Busl, G. Granger, L. Gradreau, R. Sanchez, A. Kam,
M. Pioro-Ladriers, S. A. Studenikin, P. Zawadzki, Z. R. Wasilewski,
A. S. Sachrajda and G. Platero, Nature Nanotech \textbf{8}, 261
(2013).

\bibitem[30]{Bra} F. R. Braakman, P. Barthelemy, C. Reichi, W.
Wegscheider, and L. M. K. Vandersypen, Nature Nanotech \textbf{8},
432 (2013).

\bibitem[31]{Ama} S. Amaha, W. Izumida, S. Teraoka, S. Tarucha, J.
A. Gupta and D. G. Austing, Phys. Rev. Lett. \textbf{110}, 016803
(2013).

\bibitem[32]{Sanc} R. Sanchez, F. Gallego-Marcos and G. Platerro,
Phys. Rev. B \textbf{89}, 161402(R) (2014).

\bibitem[33]{Gue} C. M. Guedon, H. Valkenier, T. Markussen, K. S. Thygesen
, J. C. Hummelen and S. van der Molen, Nature nanotechnology
\textbf{7}, 304  (2012).

\bibitem[34]{Been} C. W. J. Beenakker and A. A. M. Staring, Phys.
Rev . B 46, 9667 (1992).

\bibitem[35]{Tseng} Y. C. Tseng, David M T Kuo, Y. C. Chang and C.
W. Tsai, arXiv:1504.06082v1

















\end{thebibliography}
\end{document}